\documentclass[journal]{IEEEtran}
\usepackage{cite}
\usepackage{amsmath,amssymb,amsfonts}
\usepackage{algorithmic}
\usepackage{graphicx}
\usepackage{textcomp}
\usepackage{caption}
\usepackage{subcaption}
\usepackage{float}
\usepackage[utf8]{inputenc}
\usepackage{bm}
\usepackage{color}
\usepackage{verbatim}
\usepackage{cleveref}
\usepackage{url}
\usepackage{balance}
\usepackage{enumitem}

\usepackage{amsfonts}
\usepackage{amsmath}
\usepackage{amssymb}
\usepackage{commath}
\usepackage{cite}
\usepackage{dsfont}
\usepackage{latexsym}
\usepackage{times}
\usepackage{url}
\usepackage{verbatim}

\def\Nt{N_{\sf t}}

\def\e0{{\epsilon_0}}

\def\Nt{N_{\sf t}}

\def\Nrf{N_{\mathsf{rf}}}

\begin{document}

%\title{{The Tri-hybrid MIMO Architecture}}
\title{{\fontsize{24pt}{28pt}\selectfont The Tri-Hybrid MIMO Architecture}}
\author{{Robert W. Heath, Jr.}~\IEEEmembership{Fellow,~IEEE},
Joseph Carlson,~\IEEEmembership{Student Member,~IEEE}, \\
Nitish Vikas Deshpande,~\IEEEmembership{Student Member,~IEEE},  
Miguel Rodrigo Castellanos,~\IEEEmembership{Member,~IEEE},\\ 
Mohamed Akrout,~\IEEEmembership{Member,~IEEE},  and
Chan-Byoung Chae,~\IEEEmembership{Fellow,~IEEE} %  \\ 
\thanks{
R. W. Heath, Jr.$^{*}$, N. V. Deshpande and J. Carlson are with the Department of Electrical and Computer Engineering, University of California, San Diego, La Jolla, CA 92093, USA.
M. R. Castellanos and M. Akrout are with the Department of Electrical Engineering and Computer Science, University of Tennessee, Knoxville, TN 37996, USA.
C.-B. Chae$^{*}$ is with the School of Integrated Technology, Yonsei University, Seoul 03722, South Korea.  *Corresponding authors. }
}

%The IEEE Communications Magazine does not charge authors for extra pages. However, the maximum number of pages of any paper is SEVEN (7) magazine pages. Manuscripts MUST always be submitted in two-column format, complete with authors’ bios. It is recommended that the manuscript be written in Times New Roman, 12-pt font size, and 20-pica column width, but authors may use an IEEE approved two-column template. The following parameters are required to meet the page limit:
%Manuscript should not exceed 5500 words in total; inclusive of title, authors’ names/info, abstract, body (introduction to conclusion), figures, tables, captions, footnotes, acknowledgements, references, authors’ bios, etc. (everything). Footnotes are strongly discouraged however. MS Word enables authors to monitor their word count easily, and is therefore recommended
%Authors’ bios are not to exceed 150 words each.
%Regardless of typesetting, the authors are responsible for maintaining a maximum of SEVEN (7) pages upon initial submission (recommended to have less than 7 pages in order to leave room for potential improvements in revisions that require additional space), revisions, and final manuscript. Manuscripts may be rejected if not complying with this limit at any time. Final, camera-ready, versions will not be published if they exceed this limit.

\markboth{}%
{Heath \MakeLowercase{\textit{et al.}}: The Tri-hybrid MIMO Architecture}

\maketitle
%\begin{center}
%    \vspace{-3.5em} 
%    \textit{Invited Paper}
%    \vspace{1.5em}
%\end{center}

% paragraph budget targeting 25 pagraphs
% 1 	Abstract
% 5  Intro
% 5  Architectures
% 4 Future directions

%%%%%%%%%%%%%%%%%%%%%%%%%%%%%%%%%%%%%%%%%%%%%%%%%%%%%%%%%%%%%%%%%
\begin{abstract}
% Below is the abstract from the journal paper
%Multiple-input multiple-output (MIMO) communication has led to immense enhancements in data rates and efficient spectrum management. The evolution of MIMO, though, has been accompanied by increased hardware complexity and array sizes, causing the system power consumption to increase. Despite past advances in power-efficient hybrid architectures, new solutions are needed to enable extremely large-scale MIMO deployments for 6G and beyond. In this paper, we introduce a novel architecture that integrates low-power reconfigurable antennas with both digital and analog precoding. This \emph{tri-hybrid} approach addresses key limitations in traditional and hybrid MIMO systems by improving power consumption and adds a new layer for signal processing. We provide an analysis of the proposed architecture and compare its performance with existing solutions, including fully-digital and hybrid MIMO systems. The results demonstrate significant improvements in energy efficiency, highlighting the potential of the tri-hybrid system to meet the growing demands of future wireless networks. We conclude the paper with a summary of design and implementation challenges, including the need for technological advancements in reconfigurable array hardware and tunable antenna parameters.

We present an evolution of multiple-input multiple-output (MIMO) wireless communications known as the tri-hybrid MIMO architecture. In this framework, the traditional operations of linear precoding at the transmitter are distributed across digital beamforming, analog beamforming, and reconfigurable antennas. Compared with the hybrid MIMO architecture, which combines digital and analog beamforming, the tri-hybrid approach introduces a third layer of electromagnetic beamforming through antenna reconfigurability. This added layer offers a pathway to scale MIMO spatial dimensions, important for 6G systems operating in centimeter-wave bands, where the tension between larger bandwidths and infrastructure reuse necessitates ultra-large antenna arrays. We introduce the key features of the tri-hybrid architecture by (i)~reviewing the benefits and challenges of communicating with reconfigurable antennas, (ii)~examining tradeoffs between spectral and energy efficiency enabled by reconfigurability, and (iii)~exploring configuration challenges across the three layers. Overall, the tri-hybrid MIMO architecture offers a new approach for integrating emerging antenna technologies in the MIMO precoding framework.  
\end{abstract}
%%%%%%%%%%%%%%%%%%%%%%%%%%%%%%%%%%%%%%%%%%%%%%%%%%%%%%%%%%%%%%%%%

%%%%%%%%%%%%%%%%%%%%%%%%%%%%%%%%%%%%%%%%%%%%%%%%%%%%%%%%%%%%%%%%%
\section{Introduction}
%%%%%%%%%%%%%%%%%%%%%%%%%%%%%%%%%%%%%%%%%%%%%%%%%%%%%%%%%%%%%%%%%

\IEEEPARstart{O}ver the past 30 years, multiple-input multiple-output (MIMO) technology has fundamentally transformed wireless communications. By employing multiple transmit and receive antennas, MIMO enables the simultaneous transmission of multiple data streams, dramatically increasing spectral efficiency. Commercial systems have evolved from basic $2 \times 2$ configurations to large-scale antenna arrays, giving rise to several MIMO paradigms: single-user MIMO for point-to-point links, multi-user MIMO for serving multiple users concurrently, and massive MIMO for realizing channel hardening and favorable propagation through large-scale spatial diversity. Each phase brought substantial gains in data rate, reliability, and spectrum utilization.

The evolution of MIMO can be broadly categorized into three phases. The first phase focused on increasing the number of antennas at both the transmitter and receiver to enhance spatial multiplexing and spectral efficiency in point-to-point links. The second phase emphasized scaling antenna arrays at the base station, where larger physical space allowed for multi-user and massive MIMO, enabling simultaneous service to many users. The third phase emerged alongside the shift to millimeter-wave (mmWave) frequencies, where shrinking wavelengths permitted (and required) denser arrays. Hybrid analog-digital architectures were introduced to manage the power and hardware complexity of large arrays at these higher frequencies~\cite{HeathEtAlOverviewSignalProcessingTechniques2016}. Each transition was driven by the dual pressures of rising data demands and the technical challenges associated with expanding bandwidth and frequency.

Today, the field is entering a new era, often referred to as extreme or giga-MIMO, characterized by ultra-large arrays operating in the upper midbands, also known as centimeter-wave (cmWave) frequencies, typically spanning $6$ to $24$ GHz~\cite{KangWireless2024}. These bands offer a compelling trade-off: they provide significantly more bandwidth than traditional sub-6~GHz frequencies, while offering better coverage and penetration characteristics than mmWave bands. As a result, there is growing interest in using this spectrum to increase capacity without too many changes to the current infrastructure.

\begin{figure*}[!t]
    \centering
    \includegraphics[width=\textwidth]{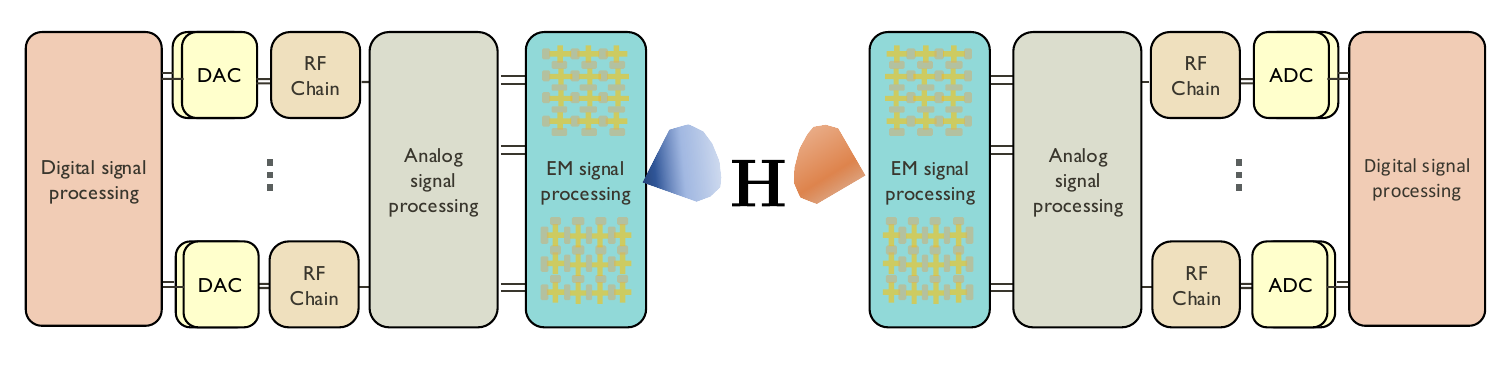}
    \caption{A general illustration of the tri-hybrid MIMO architecture. The antennas of the hybrid MIMO architecture are replaced by the new electromagnetic precoding layer that is created through the use of one or more reconfigurable antennas.}
    \label{fig_tri_hybrid_MIMO}
\end{figure*}

Operators aim to reuse the deployment density of sub-6 GHz systems, avoiding the densification typically required at higher frequencies. To meet this goal, larger arrays, potentially consisting of thousands of antenna elements, are needed to achieve sufficient aperture and array gain to support the higher carrier frequencies and wider bandwidths. The primary challenge is to implement such large arrays in an \emph{energy}- and \emph{cost-efficient} way, both from a hardware and a signal processing standpoint~\cite{Heath_switch_2016}.

In this article, we introduce the tri-hybrid MIMO architecture, illustrated in Fig.~\ref{fig_tri_hybrid_MIMO}, as a potential cornerstone for the next era of MIMO. This architecture combines three layers of spatial processing: digital beamforming, analog beamforming, and reconfigurable electromagnetic antennas. While the first two layers are well-established within the hybrid MIMO paradigm, the third layer--electromagnetic beamforming via reconfigurable antennas adds an additional degree of control. These antennas, including dynamic metasurfaces, can modify their polarization, operating frequency, and radiation patterns with minimal energy consumption \cite{Dardari_3DSP_2024}, providing increased flexibility in array design without proportional increases in RF complexity or power usage.

%\cite{Dardari_3DSP_2024} capable of modifying radiation patterns with minimal energy consumption. This added flexibility opens new design possibilities for scaling antenna arrays without the proportional increase in RF complexity or power. 

We outline the operating principles and defining features of the tri-hybrid architecture. We then examine the fundamental tradeoffs between spectral efficiency and energy consumption and discuss how both model-driven and data-driven approaches can be used to configure tri-hybrid arrays. This architectural concept is still in its early stages, offering a rich landscape for future research in hardware design, algorithm development, and performance analysis.

%We describe the key features and the operation of the tri-hybrid architecture. We then explain the spectral efficiency and energy efficiency tradeoffs that can be made. Finally, we describe some of the unique features of configuring tri-hybrid MIMO arrays using model-driven or data-driven techniques. The development of the tri-hybrid MIMO architecture has just begun, which leaves many opportunities for research related to hardware, analysis, and algorithms. 

 % In this paper, we provide an overview of the tri-hybrid MIMO architecture
%
%- brings together three layers: digital beamforming, analog beamforming and reconfigurable antennas
%
%- benefits derive from the ability of the antennas to be reconfigured at lower energy cost than active digital or analog solutions
%
%- The purpose of this article is to introduce the key features of the tri-hybrid architecture and highlight directions for future research.  

The tri-hybrid MIMO architecture unifies digital, analog, and electromagnetic beamforming into a single framework. In our early work, we introduced a prototype using dynamic metasurface antennas~\cite{CastellanosEtAlEnergyefficientTrihybridPrecodingdynamic2023}. Other innovative antenna paradigms—such as pixel antennas~\cite{Pixel_2016}, RF lens antennas~\cite{EM_lens_2014}, fluid antennas~\cite{new_fluid_2026}, and movable antennas~\cite{Zhang_2024}—can be viewed as special cases within this broader tri-hybrid framework. As such, the tri-hybrid MIMO architecture offers a promising foundation for the next generation of scalable, efficient, and adaptive wireless systems.

%The MIMO tri-hybrid architecture is a union of digital beamforming, analog beamforming and reconfigurable antennas. In our initial work, we introduced a tri-hybrid MIMO architecture that featured dynamic metasurface antennas~\cite{Tri_2025_Heath}. Other recently proposed MIMO paradigms include fluid antennas~\cite{new_fluid_2026} and movable antennas~\cite{Zhang_2024}. We view the tri-hybrid architecture as a unifying framework that includes, metasurface antennas, fluid antennas, movable antennas, and other types as special cases. 

%%%%%%%%%%%%%%%%%%%%%%%%%%%%%%%%%%%%%%%%%%%%%%%%%%%%%%%%%%%%%%%%%
\section{Explaining the MIMO architectures}
%%%%%%%%%%%%%%%%%%%%%%%%%%%%%%%%%%%%%%%%%%%%%%%%%%%%%%%%%%%%%%%%%

\begin{figure*}[!t]
    \centering
    \includegraphics[width=\textwidth]{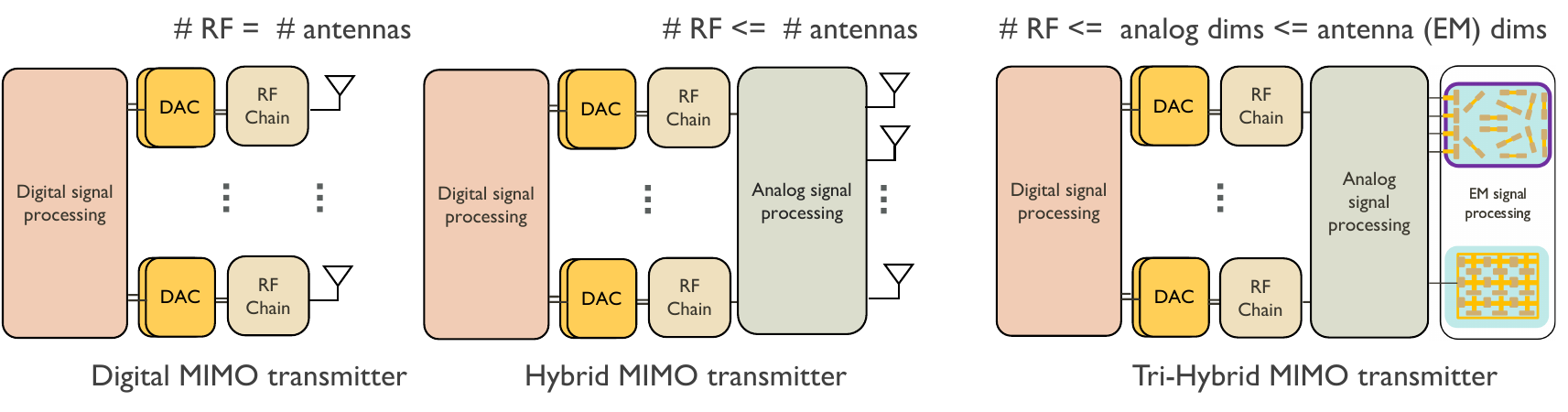}
    \caption{The digital, hybrid, and tri-hybrid MIMO architectures at the transmitter.}
\label{fig_overall_MIMO_architectures}
\end{figure*}

To provide context, we review key beamforming architectures commonly used in MIMO communication systems. For clarity,  Fig.~\ref{fig_overall_MIMO_architectures} illustrates the transmit side, though similar strategies can be used at the receiver. 

In the digital MIMO architecture, each antenna connected to a digital baseband through a radio frequency (RF) chain and dedicated mixed-signal data converter. All the signal processing like the beamforming or precoding happens in the digital domain. The power consumption is proportional to the mixed-signal (DAC in this case) and components of the RF chain including the power amplifier. The baseband precoding complexity depends on the number of RF chains. 

In the hybrid architecture, there are $\Nrf$ RF chains, each mapped to a subset of the $\Nt$ transmit antennas. In 3GPP terminology, the number of RF chains corresponds to the number of logical antenna ports, while the total number of physical antennas is referred to as the physical antenna ports. Beamforming operations in the hybrid architecture are divided between the digital and analog domains. Various connectivity options exist; for instance, in a subarray architecture, each RF chain may be connected to a specific portion of the transmit array.

%In the tri-hybrid architecture, there is a third configurable layer of antennas. The example is shown where each antenna of the hybrid architecture is replaced with a reconfigurable antenna. The reconfigurable antenna has effectively more capabilities than the static antennas, achieved for example through the interconnection of many elements. In the tri-hybrid architecture, the notion of antenna port becomes confusing. Effectively, the reconfigurable antenna has capability that is equivalent to a larger array, which we capture in the number of equivalent antenna (electromagnetic) dimensions. Multi-port reconfigurable antennas could also be envisioned, such as an adaptive dual-polarized antenna. 

In the tri-hybrid architecture, a third configurable layer is introduced at the antenna level. Specifically, each static antenna in the conventional hybrid architecture is replaced by a reconfigurable antenna. While static antennas have fixed electromagnetic properties, such as radiation pattern, polarization, or resonance frequency, reconfigurable antennas can dynamically alter these characteristics in response to control signals. This adaptability is typically achieved through tunable materials, MEMS switches, or the interconnection of multiple sub-elements within the antenna structure. As a result, reconfigurable antennas offer significantly enhanced functional degrees of freedom compared to their static counterparts. 

In this architecture, the traditional notion of an ``antenna port" becomes ambiguous, as each reconfigurable antenna can emulate the spatial behavior of a larger, virtual array. To quantify this, we introduce the concept of equivalent antenna (electromagnetic) dimensions, which captures the number of independent spatial modes that can be synthesized. Moreover, multi-port reconfigurable antennas, such as adaptive dual-polarized or pattern-diverse designs, can also be envisioned, further expanding the aperture agility and system performance.

\begin{figure}[!t]
    \centering
       \includegraphics[width=0.5\textwidth]{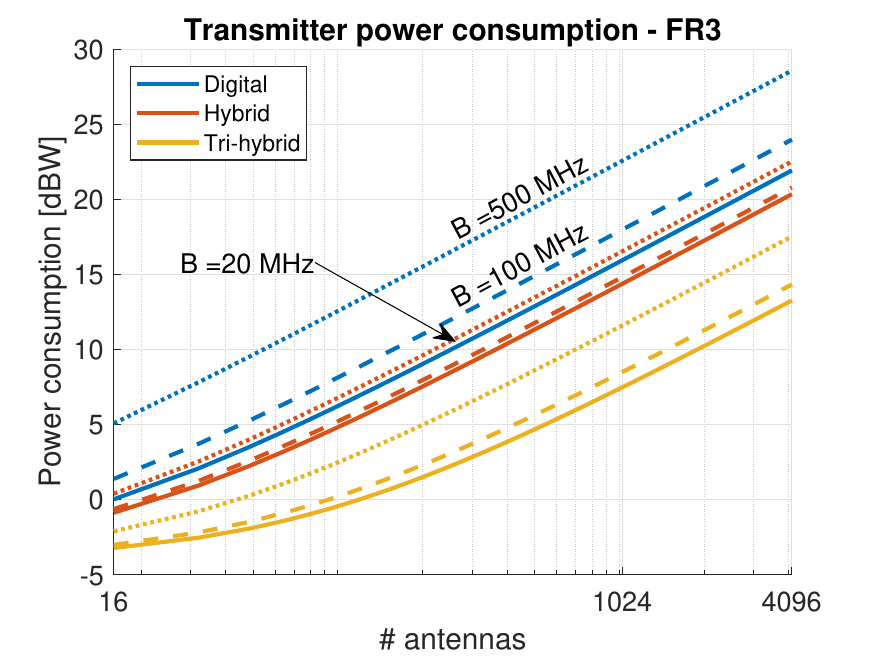}
    \caption{Power consumption for the digital, hybrid, and tri-hybrid architectures in the FR3 (upper-mid) band based on the parameters in \cite{RibeiroEtAlEnergyEfficiencyMmwaveMassive2018}. Due to the negligible power consumption of each DMA element for beamforming and the reduced number of RF chains, the tri-hybrid architecture consumes far less power than the hybrid and digital transmitter architectures.}
    \label{fig_power}
\end{figure}

We illustrate the power consumption as a function of the effective aperture size in Fig.~\ref{fig_power}. The $x$-axis refers to the number of antennas or effective antenna dimensions. The power increases with aperture as expected. Fixing a target of $10$ Watts, digital is optimal for small dimensions, hybrid for medium dimensions and tri-hybrid for large dimensions. Of course, what is not shown here is that the digital approach maintains the most flexibility and MIMO dimensions, and therefore can support the highest spectral efficiency.  Making fair comparisons remains a challenge. Further, comparisons depend a great deal on the specific type of reconfigurable antenna architecture and the split between digital, analog and reconfigurable antenna dimensions. 

There are many trade-offs to consider in the design and implementation of a tri-hybrid MIMO system. Evaluating its benefits requires well-defined metrics that capture performance from multiple perspectives, including the communication throughput achieved, the power consumed to reach that performance, and the configuration overhead associated with managing the tri-hybrid MIMO link.

%%%%%%%%%%%%%%%%%%%%%%%%%%%%%%%%%%%%%%%%%%%%%%%%%%%%%%%%%%%%%%%%%
\section{Communicating with reconfigurable antennas} \label{sec_physically_consistent_challenge}
%%%%%%%%%%%%%%%%%%%%%%%%%%%%%%%%%%%%%%%%%%%%%%%%%%%%%%%%%%%%%%%%%

%  [3-4 paragraphs](Joey's section)

\begin{figure*}[!t]
    \centering
    \begin{minipage}[b]{0.31\textwidth}
        \centering
       \includegraphics[width=\textwidth]{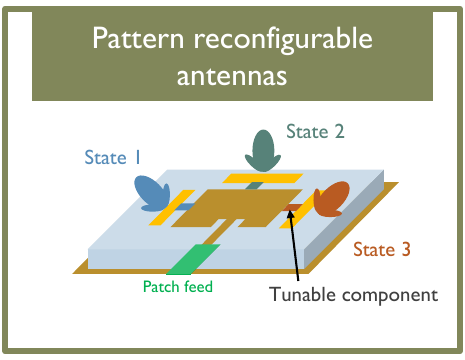}
        \subcaption{Switched-pattern antenna}
        \label{fig_switch}
    \end{minipage}
    \hspace{0.02\textwidth} 
    \begin{minipage}[b]{0.31\textwidth}
        \centering
       \includegraphics[width=\textwidth]{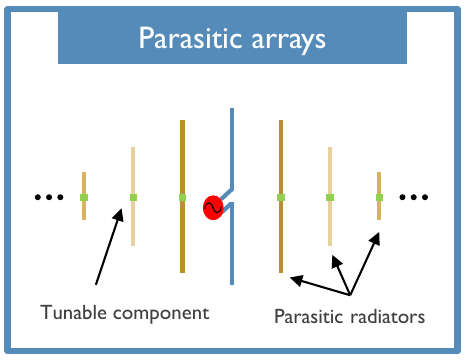}
        \subcaption{Parasitic array} %Parasitic element-assisted antenna schematic
        \label{fig_parasitic}
    \end{minipage}
    \hspace{0.02\textwidth} 
    \begin{minipage}[b]{0.31\textwidth}
        \centering
       \includegraphics[width=\textwidth]{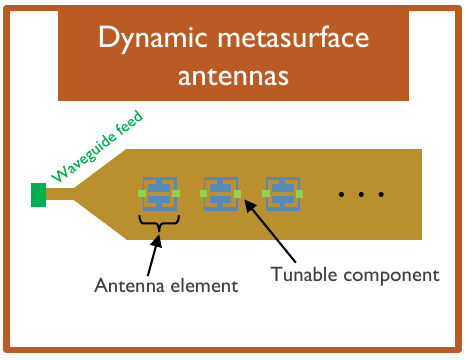}
        \subcaption{Dynamic metasurface antenna} %Structurally reconfigurable array schematic
        \label{fig_DMA}
    \end{minipage}
    \caption{Different reconfigurable antenna designs for use in the tri-hybrid architecture. Each antenna design has its own benefits and tradeoffs in terms of design complexity, insertion loss, beamforming capabilities, and consumed power. }
    \label{fig_rec_ant_types}
\end{figure*}

The performance of a tri-hybrid communication system depends on both the digital and analog connectivity, such as partially- or fully-connected hybrid architectures, as well as the type of reconfigurable antennas employed. These antennas integrate tunable components or structures to dynamically alter radiation characteristics, including pattern, polarization, or resonant frequency. Common approaches include RF switches, varactor diodes, mechanically tunable components, and dynamic metasurfaces.
%A reconfigurable antenna integrates tunable components or structures to dynamically alter its radiation characteristics. This encompasses a broad range of tunability methods and corresponding electromagnetic behaviors. For example, electrically tunable reconfigurable antennas are the most common; they typically employ RF switches or varactor diodes for reconfiguration. Other approaches include mechanically movable components or tunable substrates.

Reconfigurable antennas have been designed to adapt various parameters, including resonant frequency, polarization, element radiation pattern, and even dynamic beam steering. Given the diversity of reconfigurable antenna architectures, we now highlight and compare key design options suitable for integration into the tri-hybrid MIMO framework.

In Fig.~\ref{fig_rec_ant_types}, we illustrate three types of reconfigurable antennas. Fig.~\ref{fig_rec_ant_types}(a) shows an antenna that can reconfigure between a few different patterns. A control signal adjusts the switch state; there is loss as the signal goes through the switched path. These pattern reconfigurable antennas are typically designed such that the discrete patterns steer towards different directions to enable partial beamsteering with a single element. 

In Fig.~\ref{fig_rec_ant_types}(b), we show a parasitic array where the reactance of parasitics around an active antenna can be controlled to change its pattern. A control signal adjusts the reactance of the parasitics, similar to the pattern reconfigurable antenna. Since the array relies on mutual coupling to excite the parasitic elements, the radiation efficiency tends to deteriorate due to impedance matching. From a communication perspective, however, parasitic arrays offer a unique advantage compared to digital or analog arrays by allowing for data modulation through parasitic elements rather than the conventional RF chain. 

Fig.~\ref{fig_rec_ant_types}(c) depicts a dynamic metasurface antenna where the slots of a leaky waveguide antenna can be digitally controlled to dynamically change the radiation pattern. A control signal adjusts varactor diodes that control the resonance of the slots. DMAs are typically designed with densely-packed elements ($\sim\lambda/5$) to enable beamsteering with a reduced aperture size and power consumption \cite{Boyarsky2021}. DMAs also allow for a dynamically-reconfigurable operating frequency and bandwidth since the DMA element resonant frequencies are tunable. While all three designs in Fig.~\ref{fig_rec_ant_types} fall under the umbrella of reconfigurable antennas, each architecture offers distinct trade-offs in terms of beamforming capability, design complexity, radiation efficiency, and aperture size.
 
Performance analysis in a tri-hybrid MIMO system depends on the antenna configuration and type of reconfigurable antenna. The antenna and multipath effects are typically combined together into what is called the channel. In a tri-hybrid system, however, it becomes paramount to the channel into two distinct components: one that changes with reconfiguration and one that is only determined by the propagation environment. The former part encompasses what we call the electromagnetic precoder. This precoder maps the transmit signal from the physical antenna ports to the radiating elements constituting the entire reconfigurable array.

The dimensions and structure of the electromagnetic precoder depend on the reconfigurability and antenna design. For example, the DMA maps a single physical port to an extended array where each slot acts as a radiating element with a set of constrained weight values. Similarly for the parasitic array, the active antenna feed corresponds to the physical antenna port and the active and passive elements constitute the radiating elements. The interactions between reconfigurable antennas, through for example mutual coupling, as well as polarization effects create further challenges in analyzing the electromagnetic precoder. Due to the significant differences in reconfigurability, accurately modeling the reconfigurable antenna is a key challenge to analyzing the tri-hybrid MIMO system and determining the benefits for a broad range reconfigurable antennas.

Circuit theory offers the right tools to analyze and model the tri-hybrid MIMO architecture. This framework bridges Shannon’s mathematical information theory with electromagnetic theory by representing antennas and other RF components as multi-port networks~\cite{MezghaniEtAlReincorporatingCircuitTheory2023}. Since the radiation characteristics of realistic antenna designs are often too complex to capture analytically, except in simple cases such as the ideal dipole, the circuit-theory approach offers a simplified yet physically consistent alternative. By reducing antennas to fundamental network parameters such as impedance, admittance, hybrid, or scattering matrices, it provides a tractable model for system-level analysis.

Furthermore, the entire wireless transceiver chain, including the channel, noise sources, and impedance matching networks, can also be represented using circuit equivalents, enabling a unified model of the tri-hybrid MIMO system. While this method has been effectively applied to static antenna arrays, extending it to reconfigurable antennas requires more advanced circuit models that can capture the dynamic behaviors associated with reconfigurability and adaptive beamforming.

In parallel, full-wave EM approaches offer complementary insights, especially in capturing fine-grained phenomena such as polarization effects, mutual coupling, and near-field interactions. EM simulations and field-theoretic analyses are particularly valuable when the antenna geometry, material properties, or reconfiguration mechanisms exhibit strong spatial dependence. As such, combining EM-based modeling with circuit-theoretic abstractions could be key to fully characterizing the performance and design trade-offs of tri-hybrid MIMO systems.

% - Helps to give an asssesmsent of MIMO that is physically consistent 

% - circuit modes for antennas, circuits etc. can be used to build a general understanding of tri-hybrid MIMO.

% - need better circuit models though of reconfigurable antennas.  Ref SP magazine paper \cite{MezghaniEtAlReincorporatingCircuitTheory2023}
 
%There are many tradeoffs to be studied in the design and realization of a tri-hybrid MIMO system. Understanding the benefits requires metrics that capture the benefits from different perspectives including the potential to reconfigure the antennas energy efficiency and communication performance. This requires a good understanding how much power is really consumed in the different candidate architectures. 

%%%%%%%%%%%%%%%%%%%%%%%%%%%%%%%%%%%%%%%%%%%%%%%%%%%%%%%%%%%%%%%%%
%\section{Consumed and radiated power} 
\section{Spectral efficiency and energy efficiency tradeoff} \label{sec_interplay_challenge}
%%%%%%%%%%%%%%%%%%%%%%%%%%%%%%%%%%%%%%%%%%%%%%%%%%%%%%%%%%%%%%%%%
%  [3-4 paragraphs] (Nitish's section)

\begin{figure}[!t]
    \centering
       \includegraphics[width=0.52\textwidth]{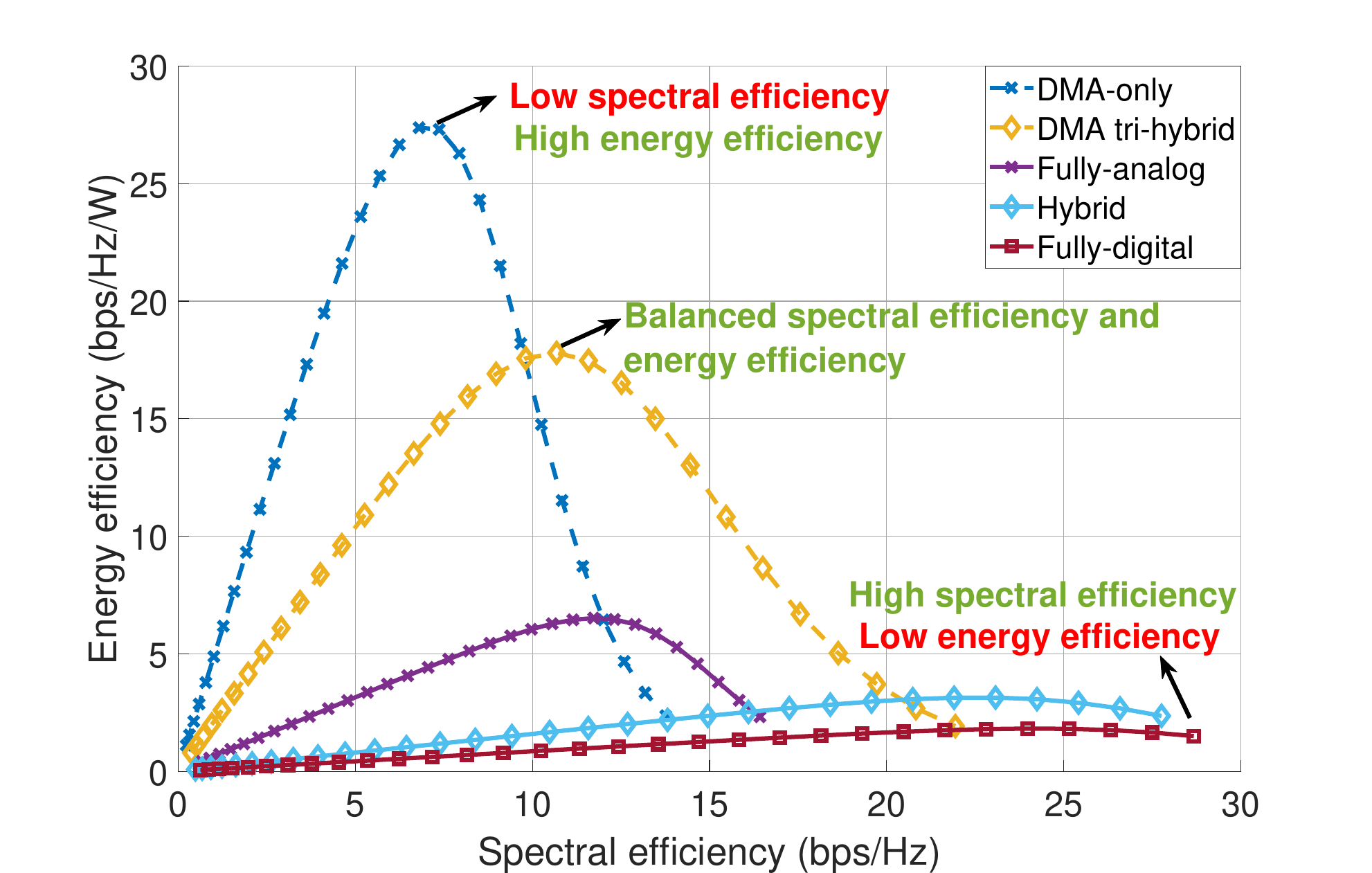}
    \caption{The tradeoff between energy efficiency and spectral efficiency is shown for different candidate architectures. The tri-hybrid DMA architecture shows a good balance between energy efficiency and spectral efficiency.}
    \label{fig_dma_ee_vs_se}
\end{figure}

% \begin{figure*}[!t]
%     \centering
%        \includegraphics[width=0.5\textwidth]{figs/consumed_vs_radiated_v1.eps}
%     \caption{Consumed and radiated power simulation??}
%     \label{fig_power}
% \end{figure*}

The benefits of the tri-hybrid MIMO architecture are especially evident from an energy efficiency perspective. Since this architecture integrates spatial processing across the digital, analog, and antenna domains, it is essential to account for losses and power consumption at each RF stage during evaluation.

While the power consumption of conventional hybrid MIMO has been well studied in the literature, several key challenges remain in performing a thorough energy efficiency analysis of tri-hybrid MIMO systems with reconfigurable antennas. The tri-hybrid framework leverages emerging types of reconfigurable antennas, such as dynamic metasurface antennas (DMAs), parasitic arrays, and polarization-reconfigurable antennas, each of which employs distinct beamforming mechanisms and therefore requires different models to accurately estimate radiated power.

%The benefits of the tri-hybrid MIMO architecture become apparent when analyzed from an energy efficiency perspective. Since the tri-hybrid architecture incorporates reconfigurability at the digital, analog, and antenna levels, it is essential to account for the losses and power consumption at each RF stage when evaluating energy efficiency. 
%While power consumption of conventional hybrid MIMO is well studied in the literature,
%%\cite{R.Mendez-RialEtAlHybridMIMOArchitecturesMillimeter2016},
%there are key challenges which need to be addressed for thorough energy efficiency analysis of tri-hybrid MIMO with reconfigurable antennas. 
%% Each reconfigurable antenna has different mechanism of beamforming which leads to different models for the received signal. 
%The tri-hybrid MIMO leverages new types of reconfigurable antennas like dynamic metasurface antennas (DMAs),
%%\cite{CarlsonEtAlHierarchicalCodebookDesigndynamic2024}, 
%parasitic antennas,
%%\cite{deshpande2025beamforming}, 
%polarization reconfigurable antennas,
%\cite{CastellanosHeathLinearPolarizationOptimizationwideband2024},
%etc, each having a different beamforming mechanism and hence different models for calculating the radiated power.

\begin{figure*}[t]
    \centering
        \includegraphics[width=\textwidth]{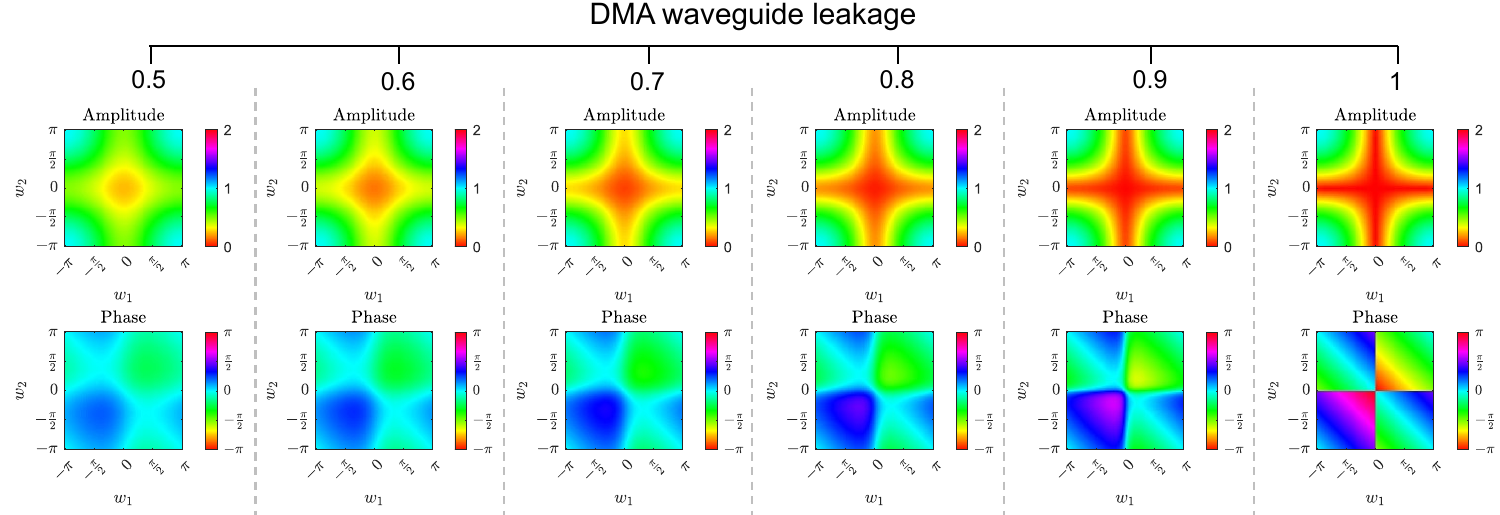}
    \caption{Magnitude and phase of the effective
transmission coefficient of a DMA with two reconfigurable elements as a function of the tunable weights $w_1$ and $w_2$ for normalized waveguide leakage values between $0.5$ and $1$. Each column corresponds to a fixed waveguide leakage value, with the amplitude shown on the top row and the phase on the bottom row.}
    \label{fig:scattering-DMA}
\end{figure*}

For example, a DMA experiences magnitude attenuation as the signal propagates through the waveguide. The loss due to attenuation must be accounted for when calculating the total energy efficiency of the DMA. Moreover, the input impedance of a DMA depends on the resonant frequency configuration of all slots, making it essential to model the effects of impedance mismatch~\cite{10942910}. In the case of hybrid parasitic arrays, the total radiated power depends on the reconfigurable reactance of the parasitic elements~\cite{deshpande2025beamforming}. As a result, the radiated power dynamically varies with each reconfiguration. Mutual coupling between the active and parasitic elements plays a critical role in determining the fraction of the supplied power that is effectively radiated from the array.

To address these challenges, a physically consistent modeling framework is essential. Examples of realistic power modeling approaches include the multi-port circuit theory framework\cite{MezghaniEtAlReincorporatingCircuitTheory2023} and EM field-based methods\cite{PizzoEtAlFourierPlaneWaveSeries2022}. In particular, power flowing into radiation modes can be tracked by modeling these modes as additional ports, enabling a more detailed, mode-by-mode analysis of power distribution, including losses arising from impedance mismatch.
% The emerging paradigm of reconfigurable parasitic hybrid arrays enables scaling to larger apertures without increasing the number of RF chains~\cite{deshpande2025beamforming}.
%@Nitish: I kept the remaining of a sentence I changed here just in case you need it: "or an EM theoretic approach based on Green's function as in []".

Beyond the losses and power consumption of devices along the signal path, the tri-hybrid MIMO architecture introduces additional auxiliary power components. Configuring the tunable elements in reconfigurable antennas typically involves low-power devices such as varactor or PIN diodes. In the case of varactor diodes, the DC biasing voltage is managed by a digital controller. While the diodes themselves draw negligible power, the digital-to-analog converter (DAC) required for tuning consumes power proportional to the resolution of control. %~\cite{TrichopoulosEtAlDesignEvaluationReconfigurableIntelligent2022}
 Consequently, the auxiliary power consumption is highly dependent on the specific hardware implementation. There exists a trade-off between the resolution of the beamforming weights and the associated auxiliary power overhead. Accurate modeling of this relationship is essential for a realistic evaluation of the energy efficiency of the tri-hybrid architecture.

The tri-hybrid MIMO architecture offers a favorable balance between energy efficiency and spectral efficiency as illustrated in Fig.~\ref{fig_dma_ee_vs_se}. While a DMA-only architecture is the most energy-efficient, it suffers from poor spectral efficiency. In contrast, fully-digital and conventional hybrid MIMO systems achieve higher spectral efficiency but at the cost of significantly increased power consumption. Tri-hybrid MIMO strikes a middle ground, delivering reasonable spectral efficiency with low power overhead, even for large-scale arrays~\cite{CastellanosEtAlEnergyefficientTrihybridPrecodingdynamic2023}. As array dimensions continue to grow, the relevance of tri-hybrid architectures becomes increasingly pronounced. Within the tri-hybrid category, multiple hardware implementations are possible. Key design factors (e.g. the type of reconfigurable antenna, array geometry, configuration mechanism, tuning resolution, and matching network) significantly influence the tradeoff between energy and spectral efficiency. A key direction for future research is identifying the optimal hardware configuration for tri-hybrid MIMO that best balances this tradeoff.

\section{Configuring the tri-hybrid array} \label{sec_configuring_challenge}
%%%%%%%%%%%%%%%%%%%%%%%%%%%%%%%%%%%%%%%%%%%%%%%%%%%%%%%%%%%%%%%%%

The reconfigurability of arrays in the tri-hybrid MIMO architecture presents significant challenges in efficiently tuning their weights (or parameters) for either beamforming or channel acquisition. This is due to the complexity introduced by their frequency, gain, and polarization adaptability, which complicates modeling, calibration, and real-time operation. Consider traditional beam training strategies, which rely on finite codebooks of predefined beam configurations: they mirror the architecture of conventional hybrid beamforming systems which employ a two-stage approach that first establishes coarse analog beam configurations before performing fine digital refinement. The proposed tri-hybrid architecture introduces an additional antenna reconfigurability layer, raising questions about modifying the reconfiguration procedure for each layer. This is particularly relevant given their distinct operational timescales, which may require new approaches to joint antenna/codebook design optimization.

%; these often perform suboptimally in dynamic environments. Because such codebooks cannot adapt to real-time channel variations or user mobility, they incur significant overhead from exhaustive or hierarchical searches and may fail to achieve accurate beam alignment.

In practice, reconfigurable elements such as varactors and PIN diodes enable dynamic beamforming through DC bias tuning in reconfigurable antennas. However, this introduces critical constraints compared to traditional phased arrays. First, finite voltage quantization limits the achievable beamforming weights, resulting in discrete phase and amplitude states that may deviate from ideal continuous values. Second, the narrow frequency tuning range of such components (e.g., the limited adaptive capacitance range of varactors) hinders wideband operation. Overall, further research is needed to explore the trade-offs between reconfigurable antenna design architectures and the resulting tri-hybrid MIMO system performance.

In addition to signal processing challenges, reconfigurable elements also exhibit  mutual coupling effects, which can distort polarization and frequency response. It can also scramble intended beam patterns to create unintended sidelobes, grating lobes, or split beams especially when these physical constraints are ignored or not properly incorporated. For example, the mutual coupling in DMAs depends on the beamforming weights and hence dynamically changes with tuned configurations. 

Fig.~\ref{fig:scattering-DMA} illustrates the trade-off in DMA design between mutual coupling and tuning sensitivity by showing the amplitude and phase of the effective transmission coefficient for a DMA with two reconfigurable elements. The coefficient characterizes the ratio of the complex scattered field to the complex feed excitation. As the normalized waveguide leakage increases from 0.5 to 1, more power is radiated earlier along the waveguide, causing the first element to receive more energy than the second. This results in an excitation imbalance between the two elements. For lower leakage values, power is distributed more evenly between the slots, and both the amplitude and phase patterns remain smooth and symmetric with respect to the tunable weights $w_1$ and $w_2$.

In Fig.~\ref{fig:scattering-DMA}, as leakage increases, the amplitude patterns become more structured, with regions of minimal scattering expanding. Simultaneously, the phase patterns become highly sensitive, with small changes in the tunable weights causing significant phase shifts in the scattered field. This highlights the complexity and sensitivity of array reconfiguration under physical constraints. When beamforming weights are optimized without accounting for mutual coupling, some elements may receive insufficient excitation while others may be unintentionally overdriven. Such an imbalance can lead to undesired sidelobe radiation or angular steering errors if mutual coupling is not properly modeled and incorporated into the design.

%Fig.~\ref{fig:scattering-DMA} illustrates the trade-off in DMA design between mutual coupling and tuning sensitivity by presenting the amplitude and phase of the scattered field attenuation for a DMA with two reconfigurable elements. As the normalized waveguide leakage increases from 0.5 to 1, more power is radiated earlier along the waveguide, resulting in the first element receiving more energy than the second. This creates an imbalance in excitation between the two elements. For lower leakage values, power is more evenly distributed, and both the amplitude and phase patterns appear smooth and symmetric with respect to the tunable weights $w_1$ and $w_2$. 

%In Fig~\ref{fig:scattering-DMA}, as the leakage increases, amplitude patterns become more structured, with regions of minimal scattering expanding. Simultaneously, phase patterns become rapidly varying, indicating that small changes in tunable weights can lead to significant shifts in field phase. This highlights the complexity and sensitivity of array reconfiguration under physical constraints. When beamforming weights are optimized without accounting for mutual coupling, some elements may receive insufficient excitation power while others may be unintentionally overdriven. This imbalance can result in undesired sidelobe radiation or angular steering errors if mutual coupling is not properly modeled and incorporated into the design.

Another important consideration is the heterogeneity of reconfigurable elements used to control the adaptive array response. Specifically, active (e.g., driven) and parasitic (e.g., load-tuned) elements exhibit fundamentally different electromagnetic (EM) behaviors. Introducing this layer of flexibility allows for non-uniform array spacing to leverage mutual coupling for enhanced performance. For instance, clustering parasitic elements near active drivers can enhance radiation efficiency or facilitate sharper nulls. Irregular element layouts can also help suppress grating lobes, an effect that is not achievable with conventional DFT-based codebooks, which assume uniform spacing and linear phase progression. In such scenarios, mutual coupling-aware algorithms are needed to jointly optimize element placement and tuning states. Viewed from this perspective, array heterogeneity becomes a valuable design degree of freedom rather than a constraint.

Reconfigurable antenna arrays offer a vast combinatorial space of possible configurations, making real-time optimization in dynamic environments a significant challenge. To address this, we see two primary approaches: model-driven and data-driven optimization. The model-driven approach relies on analytical or empirical antenna models to predict system performance and optimize configurations. Offline combinatorial methods, such as simulated annealing or genetic algorithms, can be employed when latency is not critical, while adaptive filtering techniques are suitable for real-time adaptation.

In contrast, the data-driven approach employs machine learning (ML), particularly discriminative models, to optimize configurations when the complexity of model-based methods becomes prohibitive. However, generative ML models face limitations due to their reliance on high-fidelity EM training data, the need to satisfy physics-based constraints, and the high computational cost of multi-objective EM validation. Data-driven techniques, though, provide a promising trade-off between the complexity of model-driven optimization and the stringent time requirements of real-time reconfiguration.

\section{Final thoughts}
%%%%%%%%%%%%%%%%%%%%%%%%%%%%%%%%%%%%%%%%%%%%%%%%%%%%%%%%%%%%%%%%%

%Tri-hybrid architecture offers a general MIMO framework. reconfigurable antennas offer a pathway to reducing power consumption. Significant amount of work is still needed to realize the benefits. 

The tri-hybrid MIMO architecture presents a new framework for the design of scalable and energy-efficient wireless systems, aligning well with the ambitious goals of 6G. By jointly leveraging digital and analog beamforming along with reconfigurable antennas (EM beamforming), this architecture introduces a new layer of EM control, offering unprecedented flexibility in spatial processing. While the potential benefits in terms of spectral and energy efficiency are compelling, several open issues remain before tri-hybrid MIMO can be fully realized in practice. These include:
\begin{itemize}[leftmargin=*]
    \item \textbf{Accurate and unified modeling} of diverse reconfigurable antenna architectures (e.g., parasitic arrays, dynamic metasurfaces);
    \item \textbf{Physically-consistent power and radiation modeling} across digital, analog, and EM layers;
    \item \textbf{Real-time configuration algorithms} capable of handling mutual coupling, tuning resolution constraints, and hardware nonidealities;
    \item \textbf{Trade-off analysis frameworks} that jointly consider multi-objective optimization for communication performance, hardware complexity, and control overhead.
    \item \textbf{Experimental validation} to test whether tri-hybrid models accurately characterize the behaviors of practical implementations.
\end{itemize}

Moreover, integrating these systems into realistic deployment scenarios while ensuring compatibility with existing infrastructure and protocols remains a critical challenge. As antenna technologies become increasingly heterogeneous and AI (artificial intelligence)-driven adaptation becomes essential, there is a growing need for cross-layer, cross-disciplinary research involving electromagnetics, circuit theory, signal processing, and machine learning. In this light, the tri-hybrid architecture is not just a standalone innovation but offers the potential to rethink how MIMO systems are designed, optimized, and deployed. We hope that this work encourages the community to explore these open directions and to contribute toward establishing tri-hybrid MIMO as a foundational element of the 6G wireless era.

\bibliographystyle{IEEEtran}
\bibliography{refs_rc}

\section*{Acknowledgements}
\small{This material is based upon work supported by the NSF under grant nos. NSF-ECCS-2435261, NSF-CCF-2435254, NSF ECCS-2414678  and in part by the Army Research Office under Grant W911NF2410107. This work was also supported by IITP under 6G Cloud R\&E Open Hub, and Post-MIMO (2025-RS-2024-00428780, 2021-0-00486) grants funded by the Korean government}.

\balance
% Biographies are maximum of 150 words but ideally less as this counts as part of the 7 page limit

\begin{IEEEbiographynophoto}{} \noindent
\textbf{Robert W. Heath, Jr.} (Fellow, IEEE) is  the Charles Lee Powell Chair of Wireless Communications with the Department of Electrical and Computer Engineering, University of California at San Diego, CA, USA. He is also the President and the CEO of MIMO Wireless Inc. He has authored \emph{Introduction to Wireless Digital Communication} (Prentice Hall, 2017) and \emph{Digital Wireless Communication: Physical Layer Exploration Lab Using the NI USRP} (National Technology and Science Press, 2012) and coauthored \emph{Millimeter Wave Wireless Communications} (Prentice Hall, 2014) and \emph{Foundations of MIMO Communication} (Cambridge University Press, 2018).  He received the 2025 IEEE/RSE James Clerk Maxwell Medal. He is an elected member of the National Academy of Engineering in the USA.
\end{IEEEbiographynophoto}

\begin{IEEEbiographynophoto}{}
\noindent
\textbf{Joseph Carlson} (Student Member, IEEE) received the B.S. in engineering physics from The Ohio State University, OH, USA in 2021 and the M.S. in electrical and computer engineering from NC State University, NC, USA in 2023. He is currently pursuing a Ph.D. in electrical and computer engineering at UC San Diego, CA, USA. In 2024, he interned with the antenna design team at FirstRF Corporation and received the 2023-2024 Qualcomm Innovation Fellowship. His research focuses on reconfigurable antennas, dynamic metasurface arrays, and their integration into MIMO wireless systems.
%\textbf{Joseph Carlson} (Student Member, IEEE) received the B.S. degree in engineering physics from The Ohio State University, Columbus, OH, USA, in 2021 and the M.S. degree in electrical and computer engineering from NC State University, Raleigh, NC, USA, in 2023. He is currently pursuing the Ph.D. degree with the Department of electrical and computer engineering, UC San Diego, La Jolla, CA, USA. In 2024, he interned with the antenna design team at FirstRF Corporation, Boulder, CO, and he received the Qualcomm Innovation Fellowship for 2023-2024. His research interests include reconfigurable antenna designs, dynamic metasurface antennas, and their integration into MIMO wireless communication systems.
\end{IEEEbiographynophoto}{}

\begin{IEEEbiographynophoto}{}
\noindent
\textbf{Nitish Vikas Deshpande} (Student Member, IEEE) received the B.Tech. from IIT Kanpur, India in 2021 and the M.S. from NC State University, USA in 2023. He is currently a Ph.D. student at UC San Diego, USA under Prof. Robert W. Heath Jr. He interned at Nokia Bell Labs (2022–2023), where he received the Outstanding Student Research Award, and was awarded the 2023-2024 Qualcomm Innovation Fellowship. His research interests include signal processing and optimization for advanced MIMO systems.

%\textbf{Nitish Vikas Deshpande} (Student Member, IEEE) received the B.Tech. degree in electrical engineering from the Indian Institute of Technology (IIT), Kanpur, India, in 2021, and the M.S. degree from North Carolina State University in 2023. He is currently pursuing the Ph.D. degree with the Department of Electrical and Computer Engineering, University of California San Diego, under the supervision of Prof. Robert W. Heath Jr. From 2022 to 2023, he interned with the Radio Systems Research Department, Nokia Bell Labs, where he received the Outstanding Student Research Award. He was a recipient of the Qualcomm Innovation Fellowship for 2023–2024. His current research focuses on signal processing and optimization for advanced MIMO communication systems. At IIT Kanpur, he was awarded the Proficiency Prize in Electrical Engineering for his outstanding undergraduate project in 2021.
\end{IEEEbiographynophoto}{}

\begin{IEEEbiographynophoto}{}
\noindent
\textbf{Miguel Rodrigo Castellanos} (Member, IEEE) received the B.S. (Hons.) and Ph.D. degrees in electrical engineering from Purdue University, USA in 2015 and 2020, respectively. He was a postdoctoral researcher at UT Austin and is currently an Assistant Professor at the University of Tennessee, Knoxville, USA. He has held research internships at Qualcomm and Nokia. His interests include massive MIMO, near-field communication, antenna theory, and signal processing. He is a co-recipient of the 2022 Fred W. Ellersick MILCOM Best Paper Award and the 2022 Vanu Bose Best Paper Award.
%\textbf{Miguel Rodrigo Castellanos} (Member, IEEE) received the B.S. (Hons.) and Ph.D. degrees in electrical engineering from Purdue University, West Lafayette, IN, USA, in 2015 and 2020, respectively. From 2020 to 2021, he was with The University of Texas at Austin. Since 2021, he has been with North Carolina State University and is currently an Assistant Professor in the Department of Electrical Engineering and Computer Science at the University of Tennessee, Knoxville, TN, USA. During the summers of 2017 and 2019, he was a Research Intern at Qualcomm Flarion Technologies, Bridgewater, NJ, USA, and Nokia Networks, Naperville, IL, USA. His research interests include massive MIMO, near-field communication, antenna theory, exposure-constrained communication, and signal processing. He coauthored an article that won the 2022 Fred W. Ellersick MILCOM Best Paper Award and the 2022 Vanu Bose Best Paper Award.
\end{IEEEbiographynophoto}

\begin{IEEEbiographynophoto}{}
\noindent
\textbf{Mohamed Akrout} (Member, IEEE) received the B.E. from Ecole Polytechnique de Montreal, Canada and the Diplome d'Ingenieur degree from Telecom ParisTech (ENST), France in 2016, the M.S. in AI from the University of Toronto, Canada in 2018, and the Ph.D. in electrical engineering from the University of Manitoba, Canada in 2024. He interned at Ansys and TU Munich during 2023–2024 and is currently a visiting scholar at UC San Diego, USA. His research focuses on signal processing for inverse problems and physics-aware antenna design. He received the NSERC doctoral scholarship in 2022.

%\textbf{Mohamed Akrout} (Member, IEEE) received the B.E. degree in computer engineering from Ecole Polytechnique de Montreal, Montreal, Canada, in 2016, the Diplome d'Ingenieur degree from Telecom ParisTech (ENST), Paris, France, in 2016, and the M.S. degree in artificial intelligence from the University of Toronto, Toronto, Canada, in 2018. He obtained the Ph.D. degree in electrical engineering from the University of Manitoba, Manitoba, Canada, in 2024. During the summers of 2023 and 2024, He was a research intern with Ansys and Technical University of Munich (TUM). He is currently a visiting scholar at the University of California San Diego (UCSD). His current research interests include signal processing for inverse problems and physically consistent antenna design. He was a recipient of the doctoral scholarship from the Natural Sciences and Engineering Research Council of Canada (NSERC) in 2022.
\end{IEEEbiographynophoto}{}

\begin{IEEEbiographynophoto}{} 
\noindent
\textbf{Chan-Byoung Chae} (Fellow, IEEE) is an Underwood Distinguished Professor and Lee Youn Jae Fellow (Endowed Chair Professor) with the School of Integrated Technology at Yonsei University, Seoul, South Korea. He is also the Chief Scientific Officer (CSO) of SensorView, Ltd. He is the author of \emph{Signal Processing Engineering: An Intuitive Approach} (Springer, 2026) and has co-authored more than 200 journal papers. He has received several awards from CES, IEEE Communications Society (ComSoc), IEEE Signal Processing Society (SPS), and IEEE Vehicular Technology Society (VTS). He is an elected member of the National Academy of Engineering of Korea.

\end{IEEEbiographynophoto}

\end{document}